# Tuning Optical Properties of FTO via Carbonaceous $Al_2O_3$ Microdot Deposition by DC plasma sputtering

Sarah Salah[1], Ahmed Atlam[2], Nagat Elkahwagy[1], Abdelhamid Elshaer[1], Mohammed Shihab[2,3]
[1] Department of Physics, Faculty of Science, Kafrelsheik University, Kafr-Elsheik, Egypt
[2] Department of Physics, Faculty of Science, Tanta University, Tanta 31527, Egypt
[3] Department of Physics, Basic Science, Alsalam University, Gharbia 31711, Egypt
E-mail: Mohammed.Shihab@science.tanta.edu.eg

**Abstract**

Fluorine-doped tin oxide (FTO) is a key transparent conductive oxide for photovoltaic and optoelectronic devices, yet its high reflectance limits light-trapping efficiency. This work demonstrates a simple DC plasma sputtering approach to deposit carbonaceous $Al_2O_3$ microdots on FTO under controlled Ar, $O_2$, and Ar–$O_2$ atmospheres. For plasma discharge in the normal mode, with plasma density $10^9 cm^{-3}$ and temperature of 2 eV, Volmer–Weber growth produced discrete microdots whose size and distribution were tuned by gas composition: dense, uniform dots in Ar (~0.89 μm radius), agglomerated structures in $O_2$, and intermediate morphologies in mixed atmospheres. Structural analysis confirmed $Al_2O_3$ formation with carbon incorporation, while SEM revealed morphology-driven optical behavior. UV–Vis measurements showed that Ar–$O_2$ coatings achieved the lowest reflectance (≈5–18%) across the visible range, outperforming bare FTO and other conditions. These findings establish a clear link between sputtering parameters, surface morphology, and optical performance, offering a scalable route to anti-reflective, light-trapping coatings for next-generation solar cells and optoelectronic devices.

## 1. Introduction

Transparent conductive oxides (TCOs) such as fluorine-doped tin oxide (FTO) are widely employed in photovoltaic devices due to their excellent electrical conductivity and high optical transparency in the visible spectrum, with a bandgap typically in the range of 3.6–3.8 eV. FTO is considered a robust alternative to indium tin oxide (ITO) because of its superior thermal and chemical stability, cost-effectiveness, and mechanical durability, making it suitable for solar cell architectures and optoelectronic applications [1,2,3].

Recent research has focused on surface engineering of FTO to enhance light scattering and increase the optical path length within solar cells, thereby improving absorption and charge collection efficiency. Among various strategies, the deposition of nanostructured layers or composite dots on FTO has shown promise for tailoring optical properties without compromising electrical performance [4,5].

Aluminum oxide ($Al_2O_3$) is a versatile dielectric material widely used in optical coatings due to its broad transparency range (UV to mid-IR), high refractive index (n ≈ 1.7–1.8), and excellent chemical and mechanical stability. Plasma-assisted sputtering techniques, including DC and RF magnetron sputtering, have been extensively employed to deposit $Al_2O_3$ thin films with controlled stoichiometry, density, and refractive index. Studies have demonstrated that adjusting oxygen flow and sputtering power can yield films with transmittance exceeding 95% and refractive indices between 1.50 and 1.66. Advanced methods such as pulsed DC, radiofrequency, and high-power impulse magnetron sputtering (HiPIMS) further improve film density and optical quality [6,7,8,9,10,11]. Full understanding of the plasma dynamics and nonlinearities require different theoretical models beside different experimental diagnostic techniques [12,13,14,15,16,17].

Hybrid nanostructures combining carbon and $Al_2O_3$ have attracted attention for their potential to integrate broadband absorption (from carbon) with dielectric contrast (from $Al_2O_3$), enabling unique optical and electrical functionalities. While carbon-based coatings are known for tunable absorption and conductivity, their integration with $Al_2O_3$ in dot form on FTO substrates remains underexplored. Preliminary studies on nanoparticle-enhanced TCOs indicate that particle size and distribution significantly influence light scattering and haze, which are critical for improving solar cell efficiency [18,19,20].

The optical behavior of nanodots is strongly dependent on their radius and spatial distribution. In the Mie scattering regime, where particle size is comparable to the wavelength of light, scattering intensity is governed by refractive index contrast and particle density. For carbon–$Al_2O_3$ dots on FTO, this interplay can be exploited to optimize transmission, haze, and absorption characteristics for photovoltaic and photonic applications [20,21].

Despite these advances, there is limited quantitative understanding of how carbon–$Al_2O_3$ dot morphology affects optical properties when deposited via DC plasma sputtering. This work addresses this gap by (i) depositing carbon–$Al_2O_3$ nanodots on FTO using DC plasma sputtering, (ii) characterizing dot size and distribution, and (iii) correlating these parameters with optical performance. The findings aim to establish a synthesis–structure–property relationship for this novel hybrid system, with implications for next-generation solar cells and optoelectronic devices.

## 2. DC plasma sputtering reactor

A direct current (DC) plasma reactor was employed for thin-film deposition. The reactor configuration consists of two parallel circular copper electrodes separated by an inter-electrode gap of 2.8 cm, with each electrode having a diameter of 5 cm, as shown in figure 1. The choice of copper electrodes ensures high electrical conductivity and efficient thermal dissipation during discharge operation. The upper electrode is partially covered with an Al/$Al_2O_3$ composite foil, which is mechanically secured using an acrylic castle nut to maintain structural stability. The effective active area of the upper electrode corresponds to approximately half of the Al/$Al_2O_3$ sheet, while the remaining portion is insulated using acrylic material. This asymmetric electrode configuration was intentionally implemented to modify the local electric field distribution and plasma sheath characteristics. The lower electrode functions as the substrate holder and supports a fluorine-doped tin oxide (FTO) coated glass substrate during deposition. FTO substrates were selected due to their high electrical conductivity, optical transparency, and thermal stability, which are desirable for applications in optoelectronic and photovoltaic devices. Positioning the substrate on the cathode enhances ion-assisted deposition processes, as positive ions from the plasma sheath are accelerated toward the substrate surface, thereby improving film adhesion, density, and microstructural properties.

The plasma discharge was generated at a working pressure of 1.4 mbar under an applied direct current (DC) voltage. The high-voltage power supply used in this study has a maximum output of 1.5 kV and a current limit of 30 mA. A 25 Ω series resistor was incorporated into the circuit to enable accurate current measurement during plasma operation. The selected pressure range was chosen to ensure operation within the normal glow discharge regime, which is characterized by stable plasma conditions and relatively uniform ion flux distribution across the electrode surface. The experimental procedure began by evacuating the discharge chamber using a rotary pump to reach the desired operating pressure. Subsequently, the applied voltage gradually increased until plasma breakdown occurred, typically at voltages slightly below 1 kV. Upon breakdown, the discharge current increased to approximately 5 mA, while the voltage measured between the electrodes decreased and stabilized at approximately 400 V.

Further increases in the power supply voltage resulted in an increase in the discharge current, reaching approximately 22 mA, while the inter-electrode voltage remained nearly constant at 400 V. This behavior is consistent with the characteristics of normal glow discharge, where the discharge voltage remains relatively stable and the current increase is mainly associated with expansion of the active plasma region across the cathode surface. After plasma ignition, the discharge voltage remained stable around 400 V, confirming steady-state plasma operation.

At the selected pressure and electrode spacing, the mean free path of charged particles provides favorable conditions for efficient energy transfer through electron–neutral collisions. These collisions promote ionization and excitation processes that are essential for sustaining plasma discharge. Furthermore, the chosen discharge parameters provide an appropriate balance between plasma density and ion bombardment energy, which is crucial for controlling thin-film growth rate, surface morphology, and compositional uniformity. To investigate the influence of reactive and inert plasma environments on the deposition process, different gas compositions were employed, including pure argon (Ar), pure oxygen ($O_2$), and Ar/$O_2$ gas mixtures.

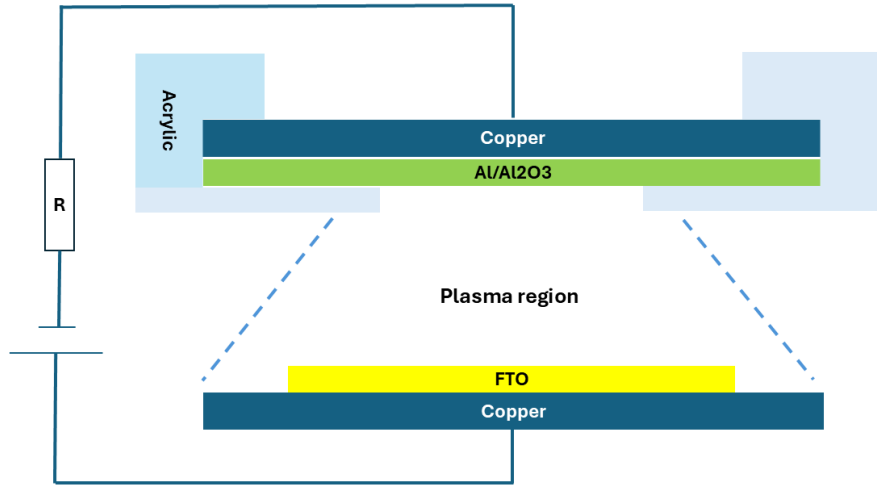

Figure (1) *Schematic representation of the DC plasma reactor used for thin-film deposition.*

## 3. Results of carbonaceous $Al_2O_3$ Microdots deposited on FTO substrate

Microdot formation was achieved through the Volmer–Weber (island) growth mode, facilitated by limited adatom mobility and a low surface energy mismatch between $Al_2O_3$ and FTO. Short deposition times and moderate oxygen partial pressure were crucial in ensuring the formation of discrete microdots rather than a continuous film. Thin films and dots were deposited using DC sputtering, a technique where a constant direct current is applied to a metallic target in a low-pressure gas atmosphere [22,23], causing atoms to be ejected from Al/Al2O3 foil and deposited onto the FTO substrate. We got a deposition pattern of microdots instead of continuous film. The pattern and the radius of the dots affect the optical properties of the FTO substrate.

### 3.1 X-ray Diffraction (XRD) Analysis.

X-ray diffraction (XRD) analysis was conducted to investigate the structural properties of the deposited $Al_2O_3$ dots on an FTO (Fluorine-doped Tin Oxide) substrate. The resulting diffractogram, shown in Figure 2, displays multiple diffraction peaks that correspond to both the FTO substrate and the deposited $Al_2O_3$ discontinuous layer. The most intense peak appears at 2θ ≈ 37.0°, which is attributed to the (200) crystallographic plane of the crystalline FTO substrate. Additional strong FTO peaks are observed at 26.6°, 33.9°, 51.8°, 61.6°, and 66.3°, which are consistent with the known tetragonal structure of $SnO_2$ (cassiterite) with fluorine doping, according to JCPDS card No. 41-1445 [2, 24, 25, 26, 27]. These peaks confirm that the FTO substrate retains its crystalline nature after the deposition process and provides a reliable base for the $Al_2O_3$ coating. Several weaker diffraction peaks are also observed in the pattern, notably near 25.7°, 33.1°, and 60.9°, which can be attributed to the deposited γ-Al₂O₃ dots. These peaks are consistent with the spinel-like structure of γ-Al₂O₃ and may correspond to reflections such as (220), (400), and (511)/(440), respectively, as reported in standard JCPDS data for γ-Al₂O₃ [28], a metastable phase commonly formed in thin films deposited at low to moderate temperatures. The relatively low intensity and broadening of these $Al_2O_3$ peaks suggest that the layer is microcrystalline or partially amorphous in nature. The absence of sharp, high-intensity peaks related to α-$Al_2O_3$ (corundum phase) further confirms that the $Al_2O_3$ film did not crystallize into its thermodynamically stable phase, which typically requires high-temperature annealing (>1000°C). Instead, the observed broad features indicate a disordered structure or the presence of a transitional phase such as γ-$Al_2O_3$ or δ-$Al_2O_3$, which is expected for low-temperature, thin-film deposition processes. As shown in figure 2, the gas discharge has no significant effect on the position of the XRD peaks, therefore, it strongly indicates that the samples share the same crystalline phase or have very similar lattice parameters. However, variations in peak intensity or width may reflect differences in crystallite size, preferred orientation, or degree of crystallinity rather than a change in phase.

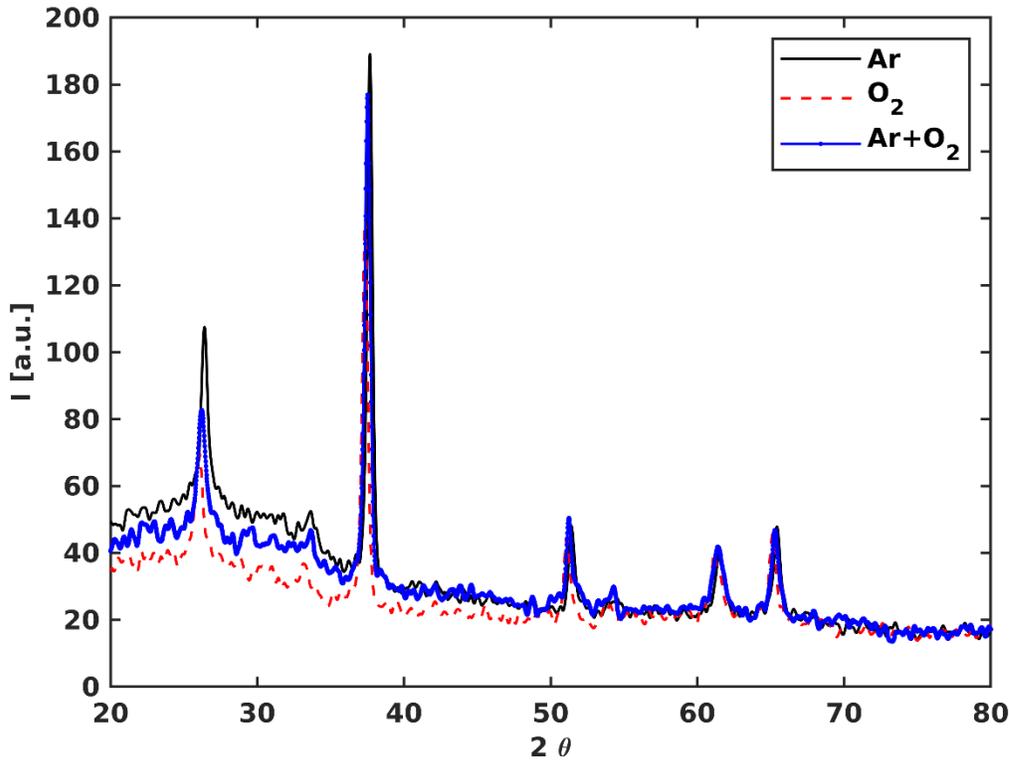

Figure (2) XRD diffraction spectrum of deposited Carbonaceous $Al_2O_3$ on an FTO (Fluorine-doped Tin Oxide) substrate using Ar, $O_2$, Ar-$O_2$ Plasmas.

### 3.2 Raman spectroscopy analysis

Following the XRD analysis, Raman spectroscopy was performed to investigate the vibrational modes and confirm the presence of various chemical bonds and functional groups in the $Al_2O_3$ dots deposited on the FTO substrate. As shown in figure 3, the Raman spectrum when the dots are deposited using Ar plasma. The Raman spectrum displays a generally low-intensity signal, which corresponds with the poor crystallinity of the $Al_2O_3$ layer indicated by the XRD results. Such weak Raman activity is typical for amorphous or poorly crystalline alumina. Several characteristic peaks were observed and assigned to different chemical species as shown in figure 4. Raman spectroscopy confirmed the presence of multiple functional groups and structural features in the carbonaceous $Al_2O_3$ dots deposited on FTO substrates. Characteristic $Al_2O_3$-related peaks, marked by purple dashed lines, were observed at low wavenumbers, indicating aluminum oxide formation despite its low crystallinity, consistent with previous reports on OH-bearing alumina phases [29,30]. Strong, yellow-marked peaks correspond to SnO vibrations from the underlying FTO substrate, in agreement with literature on fluorine-doped tin oxide [31]. Additional bands near 2000 cm$^{-1}$ (green dashed lines) suggest defect-induced or second-order vibrational modes, which have been linked to resonant defect states in oxide films (Krause et al 2025). Prominent CO and $CO_2$ signals (red dashed lines) were detected, pointing to carbon incorporation likely originating from precursor decomposition. This is plausible because the $Al_2O_3$ target was fixed using an acrylic material, which can thermally degrade during deposition and release carbon-based species, as supported by studies on Raman detection of acrylic binders and carbon residues [32]. Furthermore, an unassigned peak around 1500 cm$^{-1}$ was observed, which is likely associated with carbonaceous species or residual organic fragments from the acrylic binder, commonly attributed to C=C stretching vibrations in amorphous or graphitic carbon [33,34]. Finally, intense OH stretching vibrations in the 3200–3600 cm$^{-1}$ region (blue dashed lines) confirm the presence of surface hydroxyl groups, a common feature in $Al_2O_3$-based films [29,30]. These observations collectively validate the composite nature of the material and provide insight into its chemical environment and potential surface reactivity. These Raman results are consistent with the XRD analysis and confirm the presence of $Al_2O_3$ in the film, along with clear contributions from the FTO substrate. The detection of OH and carbon-related peaks suggests possible surface contamination or incomplete oxidation, potentially influenced by the deposition parameters or post-deposition conditions.

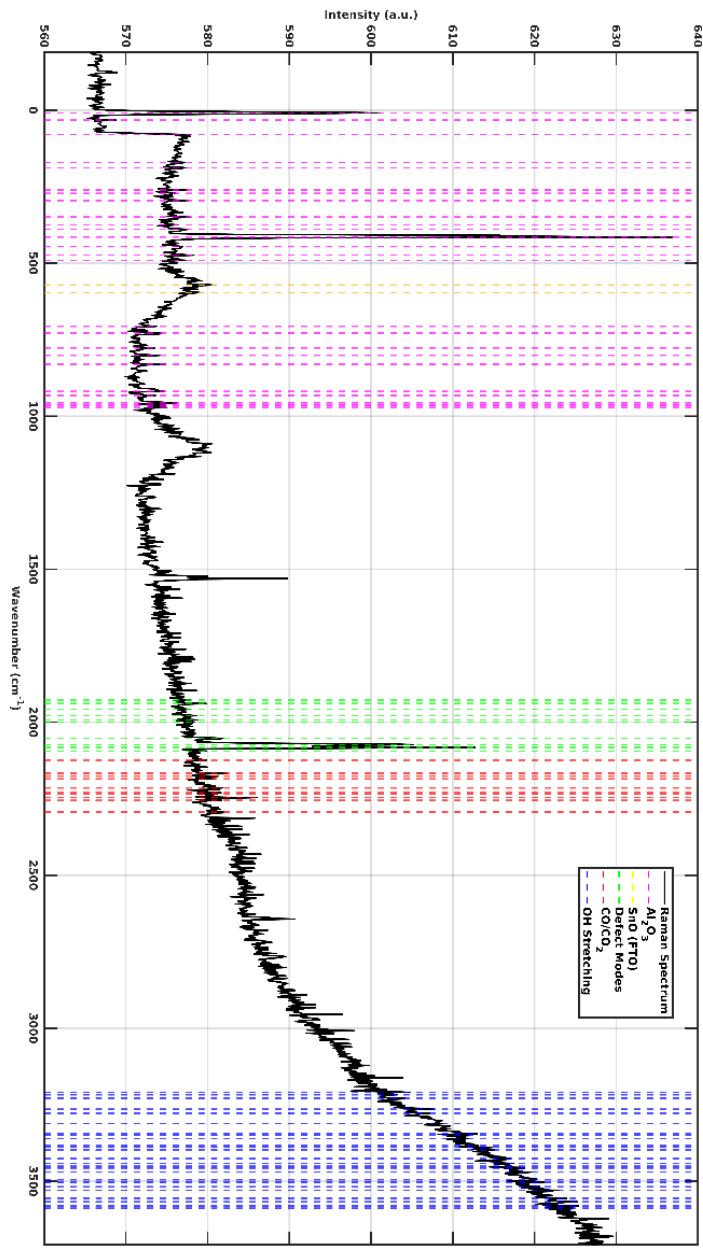

Figure (3) Raman spectrum of deposited Carbonaceous $Al_2O_3$ layer on an FTO (Fluorine-doped Tin Oxide) substrate in Ar medium.

## 3.3 Fourier Transform Infrared Spectroscopy analysis (FTIR)

Based on the FTIR spectrum, Figure 4, and the experimental setup, argon plasma discharge with a target composed of $Al_2O_3$ and an acrylic outer layer, the deposited material likely consists of a hybrid composite with the following components, explained in Table 1: The shaded regions in the FTIR absorptance spectrum correspond to characteristic vibrational modes of functional groups present in carbonaceous $Al_2O_3$ nanostructures deposited on FTO substrates. The broad band between 2800–3000 cm$^{-1}$ indicates C–H stretching vibrations, typically associated with alkyl or hydrocarbon fragments. The strong absorption near 1743 cm$^{-1}$, now included in the adjusted C=O region (1650–1800 cm$^{-1}$), is attributed to carbonyl stretching, suggesting the presence of oxidized carbon species or surface-bound organic residues. The 2100–2400 cm$^{-1}$ region represents CO/$CO_2$ asymmetric stretching, indicative of carbonate or adsorbed $CO_2$ species on the oxide surface. Peaks between 1500–1650 cm$^{-1}$ correspond to C=C stretching, commonly linked to aromatic or conjugated carbon structures, while the newly added 1400–1500 cm$^{-1}$ band highlights $CH_2$ bending vibrations, confirming alkyl chain deformation. The fingerprint region (950–1300 cm$^{-1}$) shows strong C–O stretching, associated with alcohol or ester functionalities, whereas absorptions at 700–900 cm$^{-1}$ and 400–600 cm$^{-1}$ are characteristic of Al–O and Sn–O lattice vibrations, respectively, confirming the integrity of the oxide framework. These assignments collectively demonstrate the coexistence of carbonaceous species and metal–oxygen bonds, reflecting surface functionalization and possible interactions between $Al_2O_3$ and FTO.

Table (1): Annotating peaks and functional groups in figure 4.

| Functional Group | Wavenumber Range (cm$^{-1}$) | Chemical Significance | Reference |
|---|---|---|---|
| CHx (stretching) | 2800–3000 | Alkyl C–H stretching in hydrocarbons | [35] |
| C=O (carbonyl) | 1650–1800 | Carbonyl stretching in aldehydes, ketones, esters | [36] |
| CO/$CO_2$ | 2100–2400 | Adsorbed $CO_2$ or carbonate species | [37] |
| C=C (aromatic) | 1500–1650 | Aromatic ring or conjugated double bonds | [38] |
| CH bending | 1400–1500 | $CH_2$ scissoring and $CH_3$ deformation vibrations | [39] |

| C–O | 950–1300 | Alcohol, ester, or ether C–O stretching | [40] |
| Al–O | 700–900 | Al–O lattice vibrations in alumina | [41] |
| Sn–O | 400–600 | Sn–O lattice vibrations in FTO substrate | [42] |

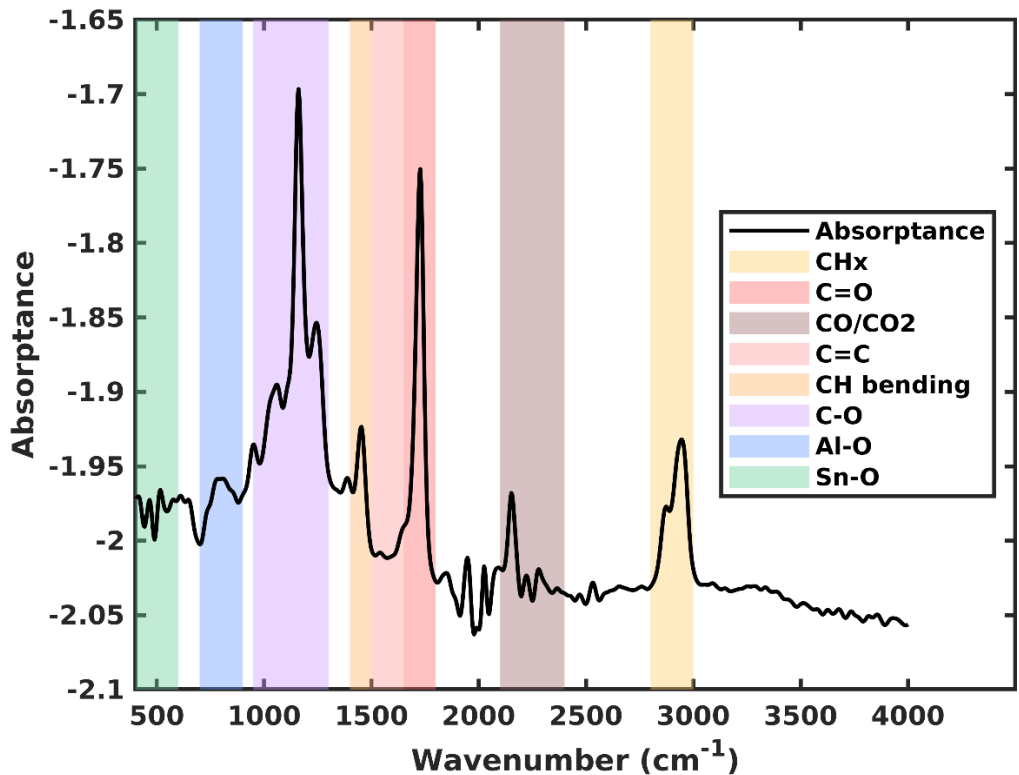

Figure (4) FTIR spectrum of carbonaceous $Al_2O_3$ microdots deposited on FTO using Ar plasma medium.

### 3.4 Scanning Electron Microscopy (SEM) and Energy-Dispersive X-ray Spectroscopy (EDS)

When carbonaceous $Al_2O_3$ is deposited on FTO substrates using DC sputtering, discrete microdots form instead of a continuous thin film, especially with short deposition times and moderate oxygen flow. This island-like morphology aligns with the Volmer–Weber growth mode, which is typically favored when the deposited species have limited surface mobility and the adatom–substrate interaction is weaker than the adatom interaction [40, 41]. Confirmed these observations, revealing microdots that were either hemispherical or irregularly rounded, with relatively sharp boundaries. Extending the deposition time of 1 hour resulted in the growth and partial merging of the microdots, indicating the beginning of percolation and film formation over longer durations. Figure 5 represents the Top-view SEM images of deposited carbonaceous $Al_2O_3$ microdots in different gas media as left panel for Ar, center panel for $O_2$, and right panel for 50% Ar+ 50% $O_2$. Each image was analyzed using a Laplacian of Gaussian (LoG) blob detection technique, calibrated using the scale bar of the image. For mixed medium (50% Ar+ 50% $O_2$) based on relative pixel size and dot size estimation, intermediate density and dot size were inferred with radius of ~ 0.6-0.7 µm and with uniformity lower than Ar but higher than $O_2$. The morphology lies between the extremes observed in pure Ar and pure $O_2$ conditions. For Ar medium, the dots were assumed circular for simplicity in radius estimation. Approximately 1411 distinct particles were detected and characterized. Average radius of dots of ~0.89 µm are estimated with a standard deviation of 0.31µm. This morphology corresponds to Volmer-Weber (island) growth mode, typical in systems where the interaction between adatoms is stronger than with the substrate. A relatively low standard deviation (~35%) indicates a homogeneous nucleation environment with minor spatial variability [41, 42]. For $O_2$ Direct SEM image inspection with scale calibration (~ 0.083 µm/pixel) was used to exclude large agglomerates (> 2 µm). Small particles were visually identified and measured. Average diameter (excluding >2 µm dots) is ~ 0.95 µm, the standard deviation is 0.20 µm and the shape of the dots is circular to slightly elliptical. In addition, the distribution of dots is irregular but distinguishable individual dots. By removing large features from analysis, a secondary population of fine nucleation sites emerges, suggesting heterogeneous nucleation. These smaller dots resemble Volmer-Weber growth but are less dense compared to pure Ar conditions. Oxygen enhances adatom mobility, leading to eventual coalescence.

Table (2): Element mapping using Energy-Dispersive X-ray Spectroscopy (EDS)

| Plasma Gas | C (mass %) | O (mass %) | Al (mass %) |
| --- | --- | --- | --- |
| Ar | ~16.8 | ~80.4 | ~2.8 |
| $O_2$ | ~23.5 | ~76.2 | ~0.3 |
| Ar–$O_2$ | ~19.8 | ~75.3 | ~4.9 |

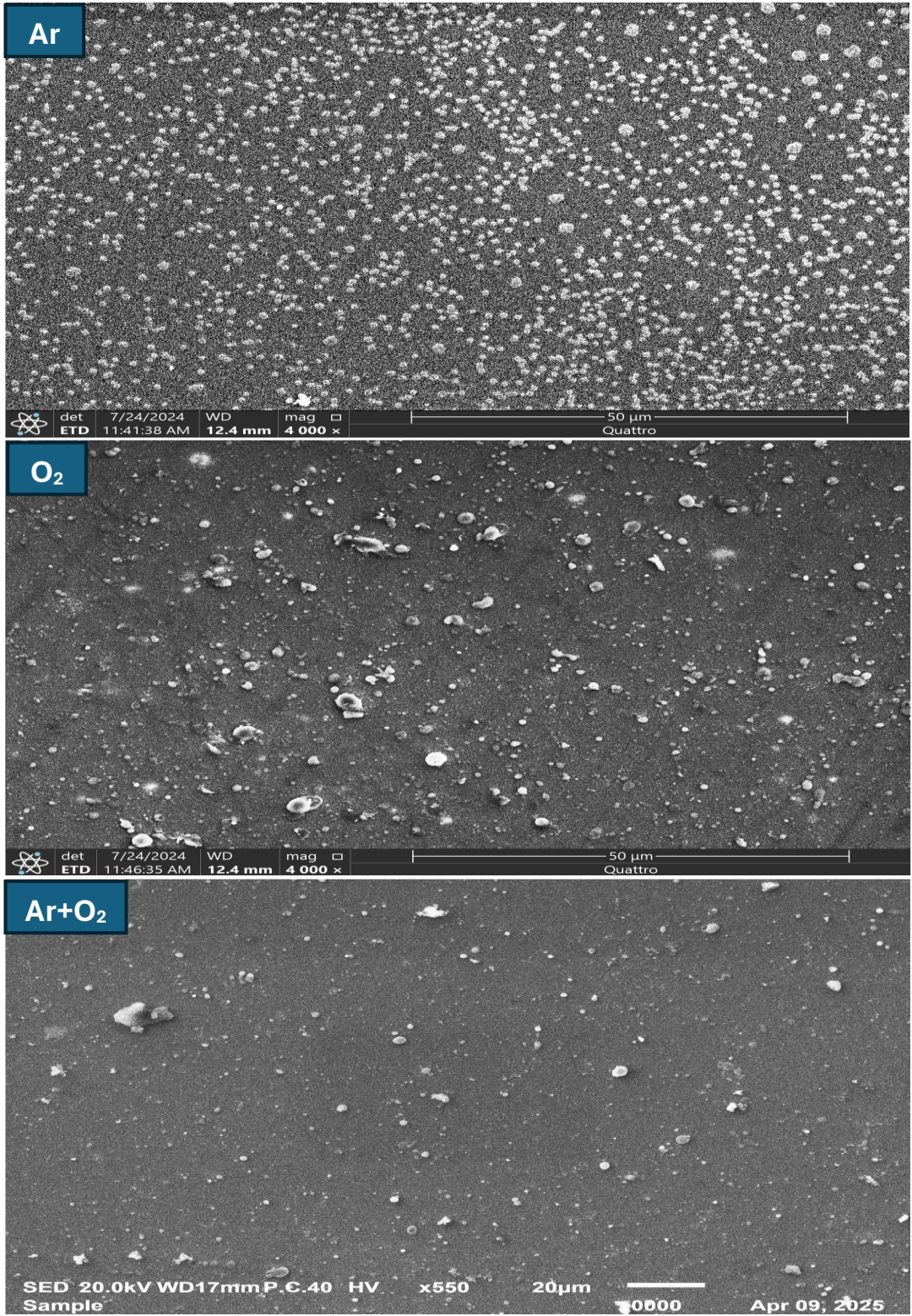

Figure (5) Top-view SEM images of carbonaceous $Al_2O_3$ Microdots deposited on FTO substrate in gas medium Ar, O2, and 50% Ar+ 50% $O_2$.

As displayed in table 2, the deposited microdots are not 2:3 of Al to O. The microdots are not pure $Al_2O_3$. However, it is $C_xAl_yO_z$, where x, y, z are fractions depending on the plasma composition. The SEM–EDS analysis confirms successful deposition of carbonaceous aluminum oxide microdots on FTO. The coating consists of oxygen-rich amorphous oxide containing dispersed $Al_2O_3$ microdomains embedded within a carbon-containing matrix. The morphology suggests uniform coverage with localized micro-agglomerates, which may enhance optical scattering and functional surface properties. The mixed Ar–$O_2$ plasma provides a synergistic sputtering–oxidation environment in which argon sustains efficient physical sputtering of the $Al_2O_3$ target, while oxygen enhances the chemical stabilization and nucleation of aluminum oxide species on the FTO substrate. This combined mechanism increases the effective sticking probability of sputtered aluminum species, resulting in higher aluminum incorporation compared with pure argon or oxygen plasmas.

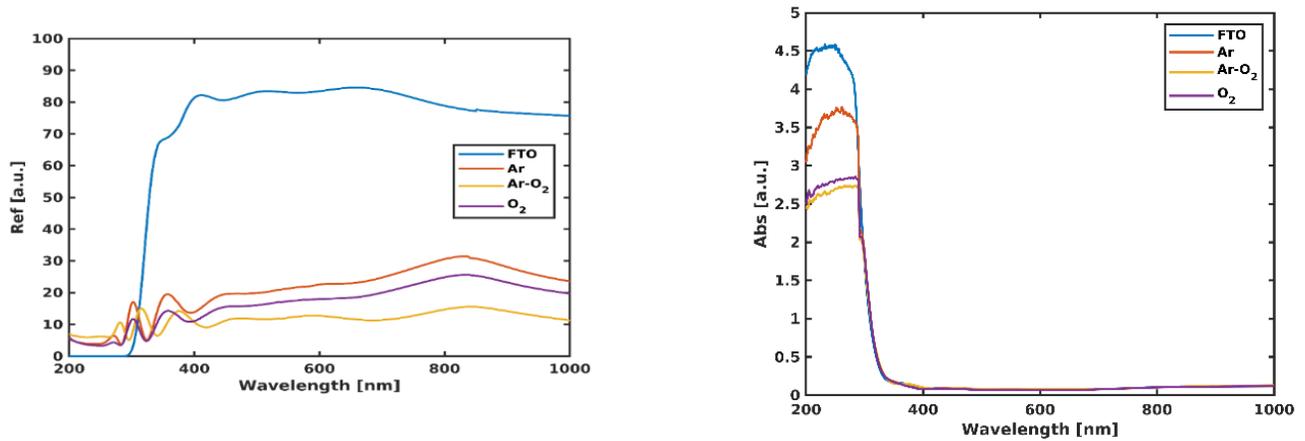

Figure (6): (Left) Represents UV-Reflectance and (right) represents the absorptance of carbonaceous $\boldsymbol{Al_2O_3}$ Microdots deposited on FTO substrate in different gas media.

## 3.5 UV-Visible Spectroscopy (UV-Vis)

Figure 6 illustrates the UV–Vis reflectance spectra of $Al_2O_3$ microdot coatings deposited on FTO substrates under different gas atmospheres, compared to bare FTO. The uncoated FTO (blue curve) exhibits high reflectance (70–85%) across the visible and near-infrared regions, consistent with its transparent and conductive nature. $Al_2O_3$ coatings significantly reduce reflectance relative to bare FTO, with distinct spectral behaviors depending on the deposition atmosphere: Ar atmosphere (orange curve): Reflectance increases gradually from ~10% to ~30% with wavelength. SEM analysis reveals a dense distribution of nanodots (~1411 per unit area) with an average radius of ~0.89 µm, indicative of Volmer–Weber (island) growth. The high density of discrete nanodots enhances surface roughness and light scattering, reducing reflectance. The wavelength-dependent increase aligns with Mie scattering from submicron particles. $O_2$ atmosphere (purple curve): Reflectance remains intermediate (~10–25%) and exhibits a smoother profile than the Ar case. SEM images show large, coalesced agglomerates (>2 µm excluded) and a heterogeneous morphology. The reduced number of sharp interfaces lowers scattering, while increased surface continuity slightly raises reflectance compared to Ar–$O_2$ but remains below bare FTO. Ar–$O_2$ atmosphere (yellow curve): Reflectance is lowest across most of the spectrum (~5–18%) with a smooth curve. SEM reveals intermediate particle density and size (~0.6–0.7 µm), representing a transitional morphology. This balance between scattering and absorption promotes light trapping, yielding the lowest reflectance—an advantageous feature for optoelectronic applications requiring anti-reflective surfaces. The UV absorption spectra (right panel in Figure 6) show a pronounced absorption edge near 320 nm for all samples, characteristic of the wide bandgap of $Al_2O_3$ and FTO. Bare FTO exhibits the highest absorption (~4.5 a.u.), which decreases significantly upon $Al_2O_3$ deposition, indicating partial coverage and modification of the optical interface. Among the coated samples, the Ar-deposited film demonstrates the highest UV absorption (~3.5 a.u.), attributed to its dense nanodot morphology formed via Volmer–Weber growth. The high density of submicron islands enhances light scattering, increasing the effective optical path length and apparent absorption in the UV region. In contrast, $O_2$-deposited films exhibit lower absorption (~2.8 a.u.), correlating with larger, coalesced agglomerates that reduce scattering and promote surface continuity. The Ar–$O_2$ atmosphere yields the lowest absorption (~2.5 a.u.), reflecting its intermediate morphology that balances scattering and coverage, consistent with its superior anti-reflective performance observed in reflectance measurements.

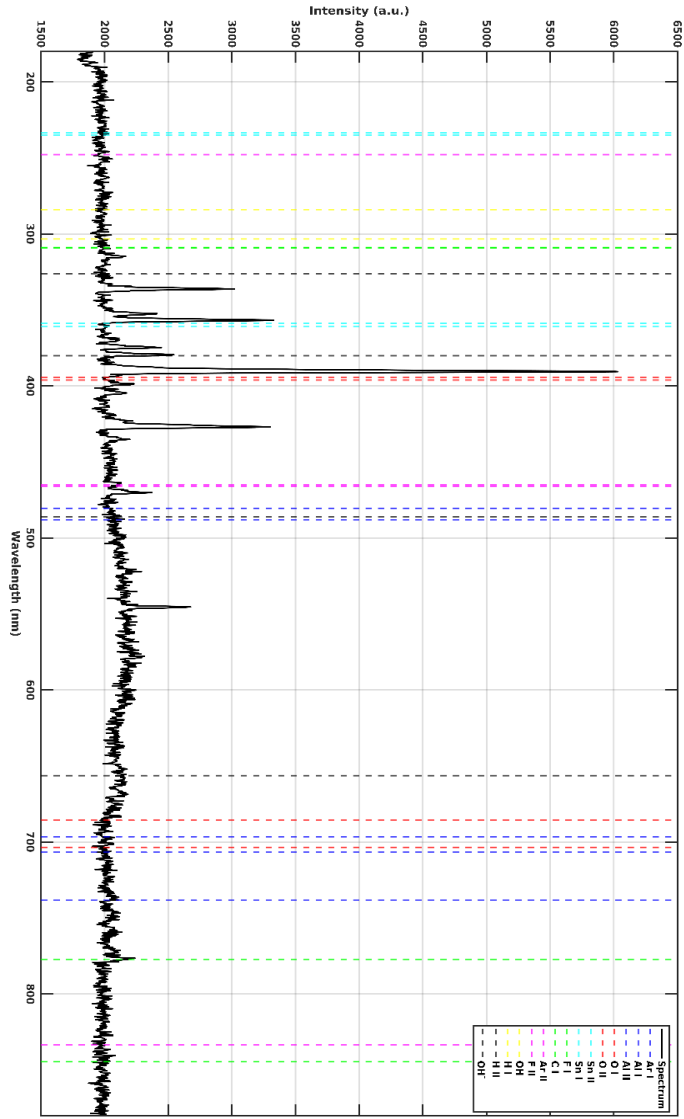

Figure (7) Compact Fiber Optic Spectrometer spectrum of Ar plasma during the deposition of Carbonaceous $Al_2O_3$ Microdots on FTO.

### 3.6 Plasma Parameters and chemical species

In addition to the structural and vibrational characterization provided by XRD, FTIR and Raman spectroscopy, Optical Emission Spectroscopy (OES) was performed to analyze the plasma conditions during the deposition of carbonaceous $Al_2O_3$ microdots. As shown in Figure 7, the emission spectrum displays multiple peaks corresponding to different species, indicating active sputtering of the target and interactions with the working gas [43]. The presence of OH and carbon-related peaks aligns with the Raman and FTIR results and suggests minor surface contamination or incomplete oxidation, likely due to deposition conditions or ambient exposure. Overall, the OES data complement the analysis, confirming the presence of $Al_2O_3$ in the deposited dots, while also highlighting contributions from the FTO substrate and possible post-deposition surface modifications.

To estimate the electron temperature of the plasma, Boltzmann plot method was employed based on the intensities of several emission lines of ArII. The graph shown in figure 8 displays the natural logarithm of the ratio Ln ($I \lambda/gA$) plotted against the excitation energy $E_k$ of the upper energy level for a series of spectral lines. Here, $I$ is the measured intensity of each line and $\lambda$ is wavelength of the emission line. Utilizing NIST Atomic Spectra Database [43] and as shown in table 3, g is the statistical weight of the upper level and A is the Einstein transition probability. The data points (blue stars) exhibit a linear trend, and a best-fit line was drawn to determine the slope, which is inversely proportional to the electron temperature according to the relation:

$$\ln\left(\frac{I\lambda}{gA}\right) = -\frac{E_k}{KT_e} + constant \qquad (1)$$

From the slope of this linear fit and Boltzmann's constant K, the electron temperature $T_e$ was calculated. The negative slope confirms a thermalized electron population, and the linearity of the data suggests the validity of the Boltzmann approximation under the experimental conditions.

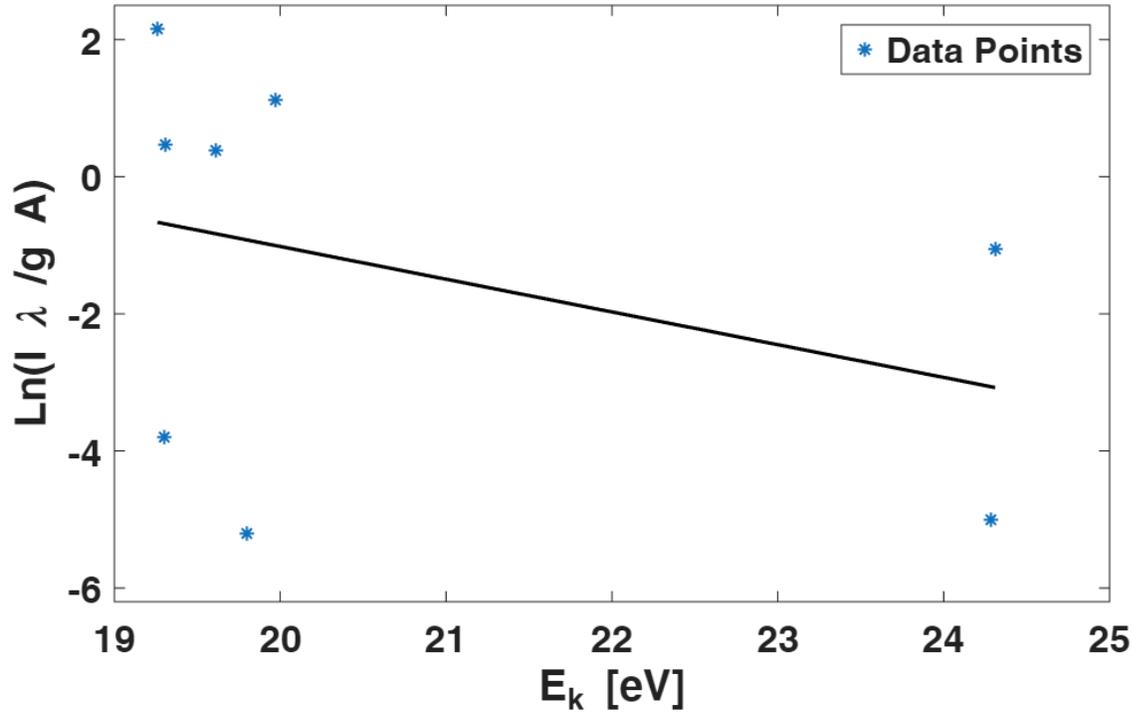

Figure (8) Boltzmann plot of Ln ($I\lambda/gA$) plotted against the excitation energy $E_k$ of the upper energy level for selected spectral lines used to estimate the electron temperature of the plasma.

Table (3): Values of parameters to generate figure 8 extracted from NIST and the measured spectrum.

| Wavelength [nm] | $E_k$ [eV] | g | A [$s^{-1}$] | I [a.u.] |
|---|---|---|---|---|
| 435.1 | 19.3 | 2 | $2.12 \times 10^7$ | 2197.76 |
| 433.7 | 24.28 | 4 | $3.4 \times 10^7$ | 2101.8 |
| 465.79 | 19.8 | 2 | $8.92 \times 10^7$ | 2118.18 |
| 470.334 | 24.31 | 4 | $8 \times 10^5$ | 2370.96 |
| 529.04 | 19.61 | 4 | $2 \times 10^5$ | 2216.49 |
| 564.24 | 19.97 | 2 | $2 \times 10^5$ | 2169.68 |
| 572.4 | 19.31 | 2 | $4 \times 10^5$ | 2221.17 |
| 584.78 | 19.26 | 4 | $3.8 \times 104$ | 2249.25 |

The electron temperature $T_e$ was found to be approximately 2 eV. These values reflect a moderately high-energy argon plasma, which is appropriate for effective sputtering and film formation. Electron density is determined using optical emission spectroscopy (OES) combined with a reduced collisional–radiative model (CRM). The model is based on population kinetics of selected argon excited states and has been widely applied for diagnosing low-temperature, low-density plasmas [44-49]. The reduced CRM considers the argon ground state ($1s_2$ in Paschen notation), two metastable states ($1s_5$ and $1s_3$), and two radiative upper levels corresponding to the Ar I emission lines at 750.4 nm ($2p_1 \rightarrow 1s_2$) and 696.5 nm ($2p_9 \rightarrow 1s_5$). These transitions were selected due to their strong emission intensity, well-characterized transition probabilities, and sensitivity to electron density variations. The model assumes steady-state conditions, a Maxwellian electron energy distribution function (EEDF), optically thin plasma, and spatially uniform plasma parameters along the line of sight. Under steady-state conditions, the populations of the radiative upper levels are governed by electron-impact excitation from the ground and metastable levels, spontaneous radiative decay, and electron-impact quenching. The population balance equations for the upper levels are expressed as $n_4 = n_e (n_0 k_{14} + n_2 k_{24}) / (A_{41} + n_e k_{q4})$, and, $n_5 = n_e (n_0 k_{15} + n_3 k_{35}) / (A_{52} + n_e k_{q5})$, where $n_e$ is the electron density, $n_0$, $n_2$, and $n_3$ represent the normalized populations of the ground and metastable states, $k_{ij}$ are electron-impact excitation rate coefficients, and $k_{qi}$ denotes electron-impact quenching coefficients. The Einstein spontaneous emission coefficients are $A_{41} = 3.8 \times 10^7$ s$^{-1}$ for the 750.4 nm transition and $A_{52} = 4.0 \times 10^7$ s$^{-1}$ for the 696.5 nm transition [46,49]. Metastable populations were assumed to be constant fractions of the ground-state population, with $n_2 = n_3 = 10^{-3}$, consistent with typical values reported for low-pressure and atmospheric argon plasmas [45,48]. The excitation rate coefficients were approximated using a Boltzmann-type expression, $k_{ij} = 10^{-8} \exp(-E_{ij} / T_e)$ cm$^3$ s$^{-1}$, where $E_{ij}$ represents the excitation energy and $T_e$ is the electron temperature in eV. Electron-impact quenching coefficients were assumed to be $k_{q4} = k_{q5} = 5 \times 10^{-8}$ cm$^3$ s$^{-1}$, introducing nonlinear density dependence in the level populations and enhancing the sensitivity of the diagnostic in the intermediate density regime. The emission intensities of the selected spectral lines were calculated from $I_{750} = n_4 A_{41}$, $I_{696} = n_5 A_{52}$, and the diagnostic line intensity ratio was defined as $R = I_{696} / I_{750}$. Model calibration curves were generated by computing R over an electron density range of $10^9$ to $10^{12}$ cm$^{-3}$ at fixed electron temperature. Electron density was then extracted by minimizing the least-squares difference between measured and modeled line ratios. This approach avoids interpolation ambiguity and remains robust even when the calibration curve exhibits non-monotonic behavior. The selected CRM formulation is particularly suitable for low-temperature argon plasmas ($T_e \approx 1$–$2$ eV), where most neutral atoms remain in the ground or metastable states and excitation processes are dominated by electron collisions. In this density range, radiative decay is the primary depopulation mechanism, while metastable states significantly enhance stepwise excitation processes. Additionally, Stark broadening of neutral argon lines is negligible compared with Doppler and instrumental broadening for electron densities below approximately $10^{12}$ cm$^{-3}$ [46,48]. Although simplified, the reduced CRM captures the dominant kinetic processes governing excited-state populations in argon plasmas and typically provides electron density estimates with uncertainties of approximately 30–50%, depending on measurement accuracy and plasma uniformity assumptions. Based on the measured spectrum and the results of Boltzmann plot, the line ratio R=0.97 and an electron temperature of 2 eV, therefore, the average electron density is $10^9$ cm$^{-3}$.

**Conclusion**

This study demonstrates a controlled approach to tuning the optical properties of fluorine-doped tin oxide (FTO) by depositing carbonaceous $Al_2O_3$ microdots via DC plasma sputtering under different gas atmospheres. Structural analyses confirmed the formation of γ-$Al_2O_3$ with carbon incorporation, while SEM revealed morphology variations governed by sputtering conditions: dense, uniform dots in Ar, agglomerated structures in $O_2$, and intermediate morphologies in Ar–$O_2$ mixtures. Optical characterization established a strong correlation between surface morphology and reflectance behavior. Among all conditions, Ar–$O_2$ deposition yielded the lowest reflectance (≈5–18%) across the visible spectrum, significantly outperforming bare FTO and other coatings. These findings highlight the potential of morphology-driven optical tuning for anti-reflective and light-trapping applications in photovoltaic and optoelectronic devices. The approach is scalable, compatible with existing sputtering systems, and offers a pathway for integrating hybrid oxide-carbon coatings into next-generation solar cell architectures. Future work should explore the electrical impact of microdot deposition and assess long-term stability under operational conditions. The optical emission spectroscopy analysis revealed an electron temperature of approximately 2 eV and an electron density on the order of $10^9$ cm$^{-3}$ under the selected discharge conditions. These plasma parameters indicate a moderately ionized glow discharge regime, providing sufficient ionization efficiency to sustain sputtering of the target materials. The relatively low electron temperature favors controlled ion bombardment energy, which influences nucleation density and growth kinetics of the deposited microstructures. Furthermore, variations in gas composition are expected to modify the electron energy distribution function, thereby affecting reactive species formation and film morphology evolution.